# Embedded Collaborative Filtering for "Cold Start" Prediction

Yubo Zhou [*, **] and Ali Nadaf[*, ***]

[*]Q.I. Leap Analytics Inc., Vancouver, BC, Canada
[**] Department of Electronic Engineering and Computer Science, York University, Toronto, ON, Canada
[***]Department of Mathematics, Simon Fraser University, Burnaby, BC, Canada

Using only implicit data, many recommender systems fail in general to provide a precise set of recommendations to users with limited interaction history. This issue is regarded as the "Cold Start" problem and is typically resolved by switching to content-based approaches where extra costly information is required. In this paper, we use a dimensionality reduction algorithm, Word2Vec (W2V), originally applied in Natural Language Processing problems under the framework of Collaborative Filtering (CF) to tackle the "Cold Start" problem using only implicit data. This combined method is named Embedded Collaborative Filtering (ECF). An experiment is conducted to determine the performance of ECF on two different implicit data sets. We show that the ECF approach outperforms other popular and state-of-the-art approaches in "Cold Start" scenarios.

## 1 Introduction

One of the main goals of a recommendation system is to understand users needs and preferences in order to suggest relevant personalized contents, products, etc. Recommendation systems influence the process of decision making for customers at large-scales and have a significant quantifiable impact on sales as well as customer experience. The improvement in the design of algorithms can lead to a more satisfactory customer experience and therefore higher conversion rates, boosting sales for various types of businesses [7].

Two dominant categories in recommendation system are *Content Based Filtering* (CBF) and *Collaborative Filtering* (CF) [17, 14]. CBF techniques use auxiliary information to predict the interest of users. However, CBF incurs extra costs for collecting auxiliary information [11]. In contrast, no auxiliary information is required for CF and it uses only user-item interactions. In case of sufficient user-item interaction data, it has been shown that CF produces more accurate results than CBF [8, 3].

Most CF techniques rely on explicit feedback of users, such as user ratings, to understand user interaction behaviour. However, collecting explicit feedback can be difficult, noisy and costly, and hence recent studies largely have restricted attention to implicit feedback data such as clicks, views, or purchases [13, 6].

One of the main challenges of recommendation systems is to enhance users engagement with the business when little valuable information is available. This situation is known as "Cold Start" scenario. Even CF, as the widely used algorithm in generating item recommendation, performs poorly in the "Cold Start" scenarios [2].

In this paper, we propose a new hybrid algorithm based on implicit feedback data when no auxiliary information is available. Items in each user session are time-independent. This paper is organized as follows: The next section provides a brief overview of the related work. In Section 3, we present details of the algorithmic components of our approach and show how the model uses CF to make recommendations. Finally, we provide an empirical examination of the results using several popular and state-of-the-art methods on both "Cold Start" and non-"Cold Start" scenarios.

## 2 Related Work

Numerous methods have been proposed to deal with the "Cold Start" and non-"Cold Start" problems. Most of them provided different approaches and claimed to provide better recommendations in their respective condition. In this section, we explain related recommendation techniques that we apply in the proposed method.

### 2.1 Item-Item K-Nearest Neighbor

K-Nearest Neighbor (KNN) is an instance-based method used for classification and regression. KNN has been widely adopted in recommendation systems [9]. The prediction is computed by applying the weighted average of $k$ nearest neighbors to approximate missing values in given target data point.

Item-Item KNN collaborative filtering is a form of CF working based on the similarity between two items purchased by a common user. This approach has been introduced by Amazon in 2003 and used since after as one of its recommender systems [10].

In this approach, a utility metric $\Phi \in \mathbb{R}^{M \times N}$ is generated from user-item interaction, where $M$ and $N$ are the number of users and items, respectively. The item vector of each item is an *M* dimensional vector as:

$$\vec{i^m} = \langle \Phi_{1,m}, \Phi_{2,m}, ..., \Phi_{k,m}, ..., \Phi_{N,m} \rangle, \quad (1)$$

where $\Phi_{k,m}$ is the entry of *k*-th row and *m*-th column of $\Phi$. To compute the similarities between two items, *l*2-norm of Cosine similarity is normally used as

$$S(i^m, i^n) = \frac{\vec{i^m} \cdot \vec{i^n}}{\left\|\vec{i^m}\right\| \cdot \left\|\vec{i^n}\right\|}. \quad (2)$$

Let $p_u$ be the user session (profile) which is composed of items purchased by user *u*. Let *I* be a set of all the item in the data-set.

For each item in the user session, the *k*-nearest items to the items and their corresponding scores (similarities) are determined using Equation (2). Then, the scores of neighboring items are aggregated, and the neighboring items with higher aggregated scores are recommended to the user. The following algorithm shows the iterative pseudo-code for this approach:

---

**Algorithm 1** Item-item collaborative filtering

1: **procedure** ITEM-ITEM-KNN
2:     rank ← ∅
3:     n ← #Neighbors
4:     k ← #Recommendation
5:     **for** *i* in $p_u$ **do**
6:       neighbors ← ∅
7:       **for** *j* in *I* **do**
8:         neighbors[j] ← S(i,j)
9:       neighbors ← sort(neighbors)[:n]
10:      **for** *j* in neighbors **do**
11:        rank[j] += neighbors[j]
12:    rank ← sort(rank)[:k]
13:    return rank

---

### 2.2 "Cold Start" Problem

"Cold Start" refers to a case when the recommendation system does not have enough information to provide the optimal possible recommendations. The "Cold Start" problem is divided into two categories: 1. Complete "Cold Start" indicates cases in which the user session contains no information. 2. Incomplete "Cold Start" occurs when the user session contains only few information [19].

Solutions for "Cold Start" problems are classified into three main classes: 1. Collecting data through interviewing "Cold Start" users [23]. In this set-up, a number of items is recommended to users in order to receive their feedback which is costly. 2. Using auxiliary information like metadata to determine the similarity between "Cold Start" users and the other users. Auxiliary information contains some user and item attributes which can be obtained from external devices such as video cameras and sensors. Hybrid approaches like a combination of CF and CBF are proposed in this category to provide an efficient recommendation [19, 8, 22]. However, due to privacy issues, collecting this information from new users is hard and high-priced. 3. Using the transaction data when no auxiliary information is available. In this situation, a powerful algorithm is required to resolve the "Cold Start" problem. The advantage of this approach is that no additional information is required to predict top recommended items. Numerous solutions such as the Singular Value Decomposition [16], the Metric Factorization [18] and rating comparison strategy (PaPare) have been proposed for this class [21]. However, the above methods compute latent factors for new users which prevent efficient real time recommendation. Using pre-built similarity model solves this problem which will be explained in Section 3.

In this study, we assume that the system recommends popular items to users in complete "Cold Start" condition. Hence, we concentrate on incomplete "Cold Start" scenarios when the data is implicit feedback. In addition, we make the assumption that no additional information is available as this is a common situation.

### 2.3 From Word Embedding to Item Embedding

Word2Vec (W2V) [12] is one of the well-known Neural Network algorithms proposed by Google in 2013 which learns the vector representations of words, called *Word Embedding*. Word embedding is a parametrized function mapping words from a sparse representation (one-hot encoding) to a dense representation. The goal of word embedding is to capture the latent similarity of words in different dimensions and distribute the word in the multi-dimensional latent feature space.

The idea of item embedding is inspired from the W2V and used in recommendation systems [1, 5]. Grbovic et al. named the modified version of word2vec, prod2vec [5]. Henceforth, for the rest of the paper, we use items and transaction history (receipts, browsing history, sessions, etc.) instead of words and context.

Two embedding models are used in W2V: Skip-Gram (SG) and Continuous Bag of Words (CBOW). These two models are similar except for the fact that in CBOW, the goal is predicting the target item from the given session, while in SG, neighboring items is predicted from the given item. The detailed explanation of the methods is given below.

**Skip Gram (SG) and Continuous Bag of Words (CBOW) models**    Let $W = \{w_1, w_2, ..., w_N\}$ be set of the items in the data-set and each item is represented by "one-hot" representation. In the CBOW, session transaction is represented by multiple items purchased in that session. The model receives a window of *c* neighboring items around the target item $w_t$ in each session which is called *window size*. The goal of the CBOW is to predict the missing item from neighboring items. The objective function used for the CBOW model is

given by

$$E = \frac{1}{D}\sum_{i=1}^{D} \log p(w_i|w_{i-c},...,w_{i-1},w_{i+1},...,w_{i+c}), \quad (3)$$

where $D$ is the number of training samples from the session transactions.

The SG attempts to learn item representation by predicting surrounding items from the given item. The SG model maximizes the average log co-occurrence probability of items that appear in training data. The objective function is given by

$$E = \frac{1}{D}\sum_{i=1}^{D} \sum_{-c \leq j \leq c, c \neq 0} \log p(w_{i+j}|w_i), \quad (4)$$

where $c$ is the window size.

The probability function $p$ for both objective functions (3) and (4) is defined by softmax function

$$p(w_o|w_i) = \frac{exp(v'^T_{w_o} v_{w_i})}{\sum_{i=1}^{N} exp(v'^T_w v_{w_i})}, \quad (5)$$

where $v_w$ and $v'_w$ are the input and output matrices. The symbol $v_{w_i}$ represents the dot product of $v_w$ and $w_i$ which is $i$-th row of $v_w$, and $v'_{w_i}$ represents the dot product of $v'_w$ and $w_i$ which is $i$-th row of $v'_w$. The goal of the training is to maximize the posterior probability $p(w_o|w_i)$ in Equation (5). The detailed description for this formula can be found in [12].

We need to compute the gradient of Equation (5) to update the input and output matrices. However, objective functions (3) and (4) are very expensive because the algorithms have to iterate through the entire item set $I$ to compute

$$\nabla \log p(v_{w_o}|v_{w_i}). \quad (6)$$

In order to be able to compute the objective functions and the gradients efficiently, a method called *Noise Contrastive Estimation* (NCE) is used to approximate the log probability. In each training step, instead of iterating through the entire items, the NCE just samples couple of negative examples from a noise distribution $P_n(W)$. The noise distribution $P_n(W)$ is computed using the ordered item frequency in the data-set. This modification changes the objective functions (3) and (4) to

$$E = -log(\sigma(v'_{w_o} \cdot h)) + \sum_{k \in W_{neg}} log(\sigma(v'_k \cdot h)), \quad (7)$$

where $\sigma(x) = \dfrac{1}{1+exp(-x)}$ is the sigmoid function, the embedded vector in the SG is given by

$$h = v_{W_I}, \quad (8)$$

the embedded vector in the CBOW is shown as

$$h = \frac{1}{2c}\sum_{i=1}^{2c} v_{w_i}, \quad (9)$$

and negative samples $W_{neg}$ are sampled from $P_n(W)$.

**Model Output** After training, the W2V algorithm drops output metric $v'_w$ and keeps input metric $v_w$ as a look up table to map $w_i$ "one-hot" representation to embedded representation $h$. Hence, the output of the algorithm is a mapping table to map $w_i$ to $h$.

## 3 Proposed Method

In this section, we describe our proposed algorithm and explain how to apply it in recommendation systems. In addition, we introduce several extensions of the method to improve the recommendation performance.

### 3.1 Item embedded representations and Recommendations

The flow chart of the ECF approach is shown in Figure 1. Items in each user session $p_u$ are expressed as embedded representation $h$ determined by Equations (8) and (9). These embedded representations of items are used in two ways for recommendation as below:

**Item-Item-KNN** We compute the neighborhood of each item in user session $p_u$ using the cosine similarity. The cosine similarities between the embedded representations of the item and its neighboring items are used as the neighborhood scores. By aggregating all neighborhood scores, a single list of neighborhood scores is created. This single list is represented as

$$[s_1, s_2, ..., s_N], \quad (10)$$

where $s_i$ indicates the scores associated to item $i$. The aggregation process is described in Algorithm 1.

**User-Item-KNN** We take the average of all embedded item vectors $h$ in user session $p_u$ to get the user vector

$$h' = \frac{1}{|p_u|} \sum_{i \in p_u} h_i, \quad (11)$$

and compute the similarity between the averaged vector $h'$ and the other embedded vector representations of items $h$. Note that the user vector $h'$ is the centroid point of items in $p_u$. Based on the similarities, the neighborhoods of the user vector $h'$ are determined.

The output of Item-Item-KNN and User-Item-KNN methods is a list of scores as shown in Equation (10). The list of recommended items is the items with higher scores.

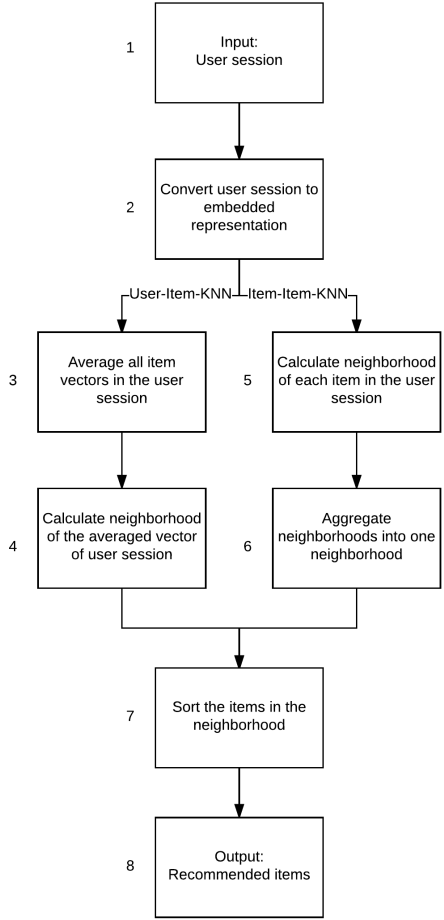

Fig. 1: Embedded Collaborative Filtering flow chart

### 3.2 Random Sampling

As already mentioned, there are no time-dependencies between items in each user session. However, the order between items matters for the W2V method. Hence, in order to destroy the arbitrary order of items appearing in user sessions and to capture all possible pair-wise correlation between items appeared in user session, we sample the items from each user session. Items in a user session are sampled randomly $m$ times, and the sampled items are used as the input of the model. The parameter $m$ is determined by

$$m = \gamma * (l - c), \qquad (12)$$

where $l$ is the length of a user session, $c$ is window size, and $\gamma$ is the random sample rate to control the sampling process. Note that the parameter $\gamma$ is chosen empirically in this work.

### 3.3 Hybrid Model

In this section, we extend our approach to improve the recommendation performance. Suppose that the individual user session is treated as a short term behaviour. Likewise, the concatenation of the individual user sessions of the same user represents the long term behavior. Accordingly, the training set is divided into two independent short- and long-term data-sets. Then, we apply the approach described in Section 2.3 separately on these two data-sets. The models generated from short term and long term data are called Short Term Model (STM) and Long Term Model (LTM), respectively. Note that the window size $c$ in STM is smaller than LTM.

The outputs of these two models are lists of scores as represented in Equation (10). The hybrid method combines the STM and LTM models as a linear combination given by

$$\Gamma = \alpha * LTM + (1 - \alpha) * STM, \qquad (13)$$

where $\alpha$ is a coefficient between 0 and 1, $\Gamma$ is the hybrid model and

$$\alpha * LTM = [s_1 * \alpha, ..., s_N * \alpha], \qquad (14)$$

and

$$(1 - \alpha) * STM = [s'_1 * (1 - \alpha), ..., s'_N * (1 - \alpha)], \qquad (15)$$

where $s_i$ and $s'_i$ are the output scores using LTM and STM, respectively. After aggregation of the scores, we rank the items based on aggregated scores in $\Gamma$ and select items with higher scores from the ranked list as the recommended items.

## 4 Experiments

We now present our setup for the experimental analysis. In our setup, we compare the proposed method with three baseline algorithms. All of these methods take transaction history as inputs and generate a list of recommended items to a given user session $p_u$. We have implemented the proposed method in Python with Gensim library [15] and executed them on a pc with intel@core i5-5200-u, 8GB DDR3L memory.

### 4.1 Data-sets

Two public stable benchmarking data-sets are used in this study as follows:

**Movielens100k**[1] is a data-set used mostly for evaluating recommendation systems. Since the data contains users ratings, it is originally explicit data. To convert it to implicit data, we keep ratings greater than 3 and change them to 1, and drop ratings less than and equal to 3.
We then create two new data-sets, short and long term data, from this implicit data. The short term data is created by grouping transactions of the same user that occurred within 7 days. Therefore, the short term data-set indicates weekly user behaviours toward movies and is used to train the STM. On the other hand, we group all transactions of the same user together as long-term data-set to train the LTM.

---

[1] https://grouplens.org/datasets/movielens/100k/

**Online Retail**[2] is a transactional data-set in a UK-based online retail store [4]. The data contains the implicit feedback and transaction time. We generate short term data by grouping transactions belong to the same user occurred in the same day. Besides that, we group all transactions belong to the same user and generate long term data. We train LTM and STM with the long term and short term data, respectively. The detailed information about the short and long term data-sets is shown in Table 1.

Table 1: Data-set statistics for short and long term data-sets

| Data-set | #Items | #Users | Density |
|---|---|---|---|
| Movielens (long term) | 1574 | 943 | 0.05556 |
| Movielens (short term) | 1574 | 1971 | 0.02659 |
| Online Retail (long term) | 3684 | 4372 | 0.01646 |
| Online Retail (short term) | 3684 | 19573 | 0.00627 |

For each data-set, we randomly hold 20% of the users in the test set and use the remaining ones in the training set. There is no overlap between users in the test and training sets. The users in the test data are also split into two portions in percentage, one portion acts as *seen items* which is used as user session $p_u$ for recommendation and the other portion is *hidden items* which is used for evaluation.

### 4.2 Evaluation Metric
**Average precision** Given the top@$N$ recommended items, the precision for each user is defined as

$$Precision@N = \frac{|R \cap H|}{N}, \quad (16)$$

where $R$ is the set of recommended items and $H$ is the set of the hidden items explained in the previous section. Note that for the top@$N$ recommendation, the size of $R$ is $N$ $\left(|R| = N\right)$.

To evaluate the performance of the algorithm, we compute the average of the precisions of all users in the test set.

### 4.3 Comparison Methods
In this section, we introduce a number of popular and state-of-the-art recommendation methods for implicit data when no auxiliary information is available. The methods are:

**POP** Items are recommended based on the popularity of items.

**CF-KNN** Items are recommended by Item-Item KNN explained in Algorithm 1.

**CDAE** [20] Collaborative Denoising Auto-Encoder (CDAE) is the state-of-the-art algorithm for top@$N$ recommendation for implicit feedback data. This approach learns a non-linear representation of items from intentionally corrupted inputs to determine the missing items in the output.

## 5 Results and Discussions
In this section, we present the comparative results of the ECF with the comparison methods based on the experiment setup described in Section 4.

### 5.1 Hyper Parameters
The main parameters of ECF algorithm include window size and random sampling rate. In this section, we illustrate a number of results associated to the model's performance for these hyper parameters. The data-set used in these experiments is the short term MovieLens data-set. These results help researchers understand the behaviour of ECF method.

**Window size** Windows size is defined as the parameter $c$ in (3). Figure 2 shows the performance of the embedding models used in ECF with different size of context windows. As observed in the figure, choices of large and small window sizes lead to worse performance. Hence, the window size should be chosen empirically for different data-sets.

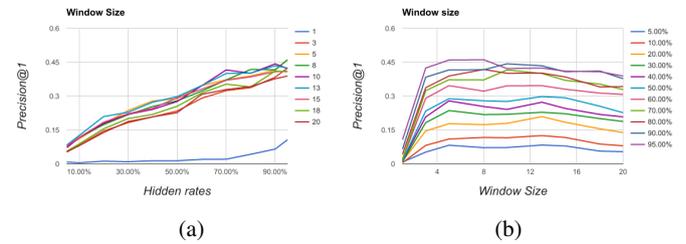

(a)      (b)

Fig. 2: Comparison of different window sizes for SG model.

**Random sampling rate** Random sampling rate in ECF has been defined in Section 3.2 as the parameter $\gamma$. Figure 3 shows the performance of ECF model for different random sampling rates. In comparison with the CBOW, the SG model requires fewer samplings from user sessions. As seen from the results, not only do the embedding models work poorly without random sampling rates, but also, random sampling with large $\gamma$ leads to worse performance.

**Item-Item-KNN vs. User-Item-KNN** Figure 4 compares the performance of two recommendation approaches of ECF, Item-Item-KNN and User-Item-KNN, described in Section

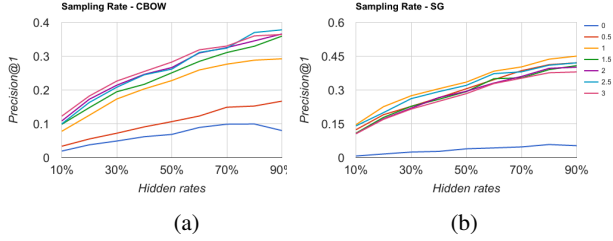

Fig. 3: Comparison of different random sampling rates between CBOW and SG models.

3.1. As observed in the figure, the User-Item-KNN approach outperforms the Item-Item-KNN.

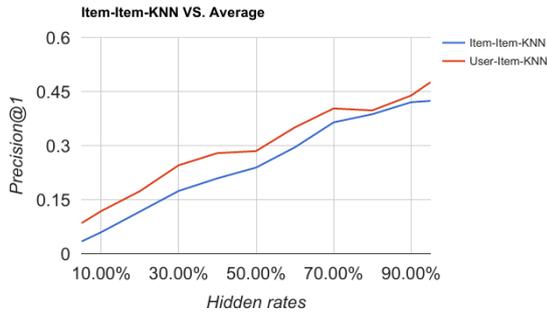

Fig. 4: Precision@1 for Item-Item-KNN vs. User-Item-KNN on short term Movielens data

**Single model vs. Hybrid model** In experiment setup, short term model (STM) is a model trained with window size 5 for each user session on short term data-set, and long term model is trained with window size 20 on long term data-set. From the experimental results shown in Figure 5, we observe that LTM performs better than STM, and the weighted combination of STM and LTM (hybrid model) leads to the optimal precision. Thus, the hybrid model learning from short and long term models, STM and LTM, provides more precise recommendations.

In future, one can find the optimal value of α for different scenarios.

**Performance Comparison** The experiment results are evaluated by precision evaluation metric (16) for top@1 and top@5 recommendations. We run 100 Monte Carlo simulations for each method to obtain more robust results. In this experiment, we use hybrid model as the embedding model for ECF. To visualize the experimental results, the precision values of ECF versus the other methods in "Cold Start" scenarios are shown in Figures 6 and 7 when 90% and 95% of items in user sessions are hidden. The experimental results in these figures show that the precisions of ECF are higher than other approaches for both data-sets in "Cold Start" scenarios.

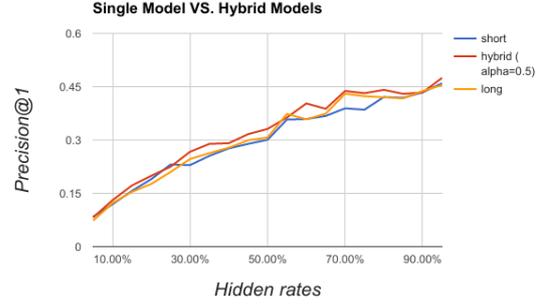

Fig. 5: Comparison between single and hybrid models

We also present the results of ECF versus the other algorithms in both "Cold Start" and non-"Cold Start" scenarios in Figure 8, where the horizontal axis represents the amount of hidden items of each user session in percentage and the vertical axis is precision. Similar to the previous figures, these figures show that ECF outperforms the other methods in "Cold Start" cases. However, in non-"Cold Start" scenarios, the system has enough information about the users, and hence, the other methods like CF or CDAE produce more accurate prediction.

Note that from the empirical results, we observed that in most cases, Embedding SG outperforms Embedding CBOW. Furthermore, the precision usually rises as the rate of hidden items increases. This is because the higher the number of hidden items, the more chance to predict them.

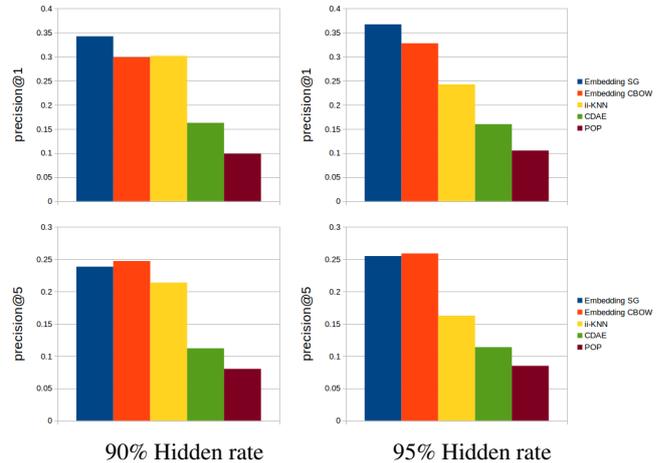

Fig. 6: The performance of different algorithms on Online retail data-set in "Cold Start" scenarios.

We show the comparison results on the MovieLens data for "Cold Start" cases in Table 2.

To summarize, the experimental results have shown that ECF obtains better accuracy in "Cold Start" scenarios than the other algorithms in terms of precision evaluation index.

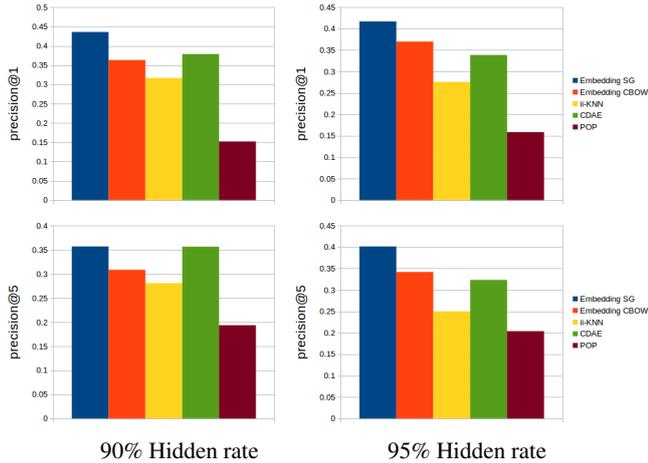

Fig. 7: The performance of different algorithms on Movielens data-set in "Cold Start" scenarios.

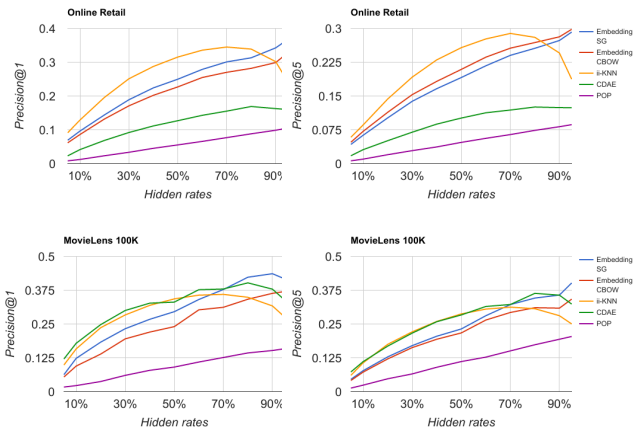

Fig. 8: The performance of different algorithms on Online retail and Movielens data-sets.

Table 2: Comparison results on MovieLens 100k data for top@1 recommendation

| Hidden rate | SG | CBOW | CF | CDAE | POP |
|---|---|---|---|---|---|
| 60% | 0.341 | 0.302 | 0.357 | **0.376** | 0.109 |
| 70% | 0.377 | 0.312 | 0.360 | **0.379** | 0.126 |
| 80% | **0.423** | 0.341 | 0.349 | 0.402 | 0.143 |
| 90% | **0.436** | 0.363 | 0.316 | 0.379 | 0.152 |
| 95% | **0.417** | 0.370 | 0.276 | 0.339 | 0.159 |

## 6 Conclusions

In this paper, we concentrated on the "Cold Start" users that negatively impact the performance of the recommender. We assumed that the data-set contains implicit feedback and that no auxiliary information is available. For this setting, we combined dimensionality reduction method with Collaborative Filtering (CF) to enhance the performance of the recommendation system in the "Cold Start" scenarios. The proposed model learns the relationship between the items in short and long term user sessions and generates a distributed representation of the items in the lower dimensional space. We used this representation of items to calculate the neighboring items for recommendation. We conducted a set of comprehensive experiments on two public data-sets to study the performance of the proposed algorithm (ECF). Through this experiment, we compared our approach with several popular and state-of-the-art algorithms in this setting and showed that our proposed approach, ECF, outperforms these algorithms in "Cold Start" scenarios.

**Acknowledgements**

The authors thank A. Akbarian, A. Rutherford, L. Fratamico, A. Wong and M. Vazifeh for fruitful discussions. This work was partially funded by the MITACS.